\def\rfr#1{Eq. (\ref{#1})}

\def\dert#1#2{\frac{{{d}}{#1}}{{{d}}{#2}}}              

\def\bb#1#2#3{\bibitem[\protect\citeauthoryear{#1}{#2}]{#3}}
\def\eqi{\begin{equation}}
\def\eqf{\end{equation}}
\def\eqia{\begin{eqnarray}}
\def\eqfa{\end{eqnarray}}
\def\rp#1#2{{#1\over#2}} \def\lb#1{\label{#1}}

\def\Om{\mathit{\Omega}}

\def\bds#1{\boldsymbol{#1}}

\documentclass{aastex}
\usepackage{url}
\usepackage{amsmath,amsthm,amscd,amssymb}
\usepackage{latexsym,wasysym}
\usepackage{graphicx,epsfig}

\RequirePackage{color}

\newcommand{\emaila}{lorenzo.iorio@libero.it}

\begin{document}

\title{General relativistic spin-orbit and spin-spin effects on the motion of  rotating particles in an external gravitational field}
\shortauthors{L. Iorio}

\author{Lorenzo Iorio\altaffilmark{1} }
\affil{Ministero dell'Istruzione, dell'Universit\`{a} e della Ricerca (M.I.U.R.)-Istruzione\\
Fellow of the Royal Astronomical Society (F.R.A.S.)\\
International Institute for Theoretical Physics and
Advanced Mathematics Einstein-Galilei\\ Permanent address for correspondence: Viale Unit\`{a} di Italia 68, 70125, Bari (BA), Italy.}

\email{\emaila}

\begin{abstract}
We analytically compute the orbital effects induced on the motion of a spinning particle  geodesically traveling around a central rotating body by the general relativistic two-body spin-spin and spin-orbit leading-order interactions. Concerning the spin-orbit term, we compute the long-term variations due to the particle's spin by finding secular precessions for the inclination $I$ of the particle's orbit, its longitude of the ascending node $\Om$ and the longitude of pericenter $\varpi$. Moreover, we generalize the well-known Lense-Thirring precessions to a generic orientation of the source's angular momentum by obtaining an entirely new effect represented by a secular precession of  $I$, and additional secular precessions of  $\Om$ and $\varpi$ as well. The spin-spin interaction is responsible of gravitational   effects \textit{\`{a}} \textit{la} Stern-Gerlach consisting of secular precessions of $I$, $\Om$, $\varpi$ and the mean anomaly $\mathcal{M}$. Such results are obtained without resorting to any approximations either in the particle's eccentricity $e$ or in its inclination $I$; moreover, no preferred orientations of both the system's angular momenta  are adopted. Their generality allows them to be applied to a variety of astronomical and astrophysical scenarios like, e.g., the Sun and its planets and the double pulsar PSR J0737-3039A/B. It turns out that the orbital precessions caused by the spin-spin and the spin-orbit perturbations due to the less massive body are below the current measurability level, especially for the solar system and the Stern-Gerlach effects.
Concerning the  solar Lense-Thirring precessions, the slight misalignment of the solar equator with respect to the ecliptic reduces the gravitomagnetic node precession of Mercury down to a $0.08$ milliarcseconds per century level with respect to the standard value of 1 milliarcseconds per century obtained by aligning the $z$ axis with the Sun's angular momentum. The new inclination precession is as large as $0.06$ milliarcseconds per century, while the perihelion's rate remains substantially unchanged, amounting to $-2$ milliarcseconds per century. Further studies may be devoted to the extrasolar planets which exhibit a rich variety of orbital and rotational configurations.
\end{abstract}


\section{Introduction}
In this paper, we deal with the issue of the dynamics of massive spinning bodies in a gravitational field in the framework of the weak-field and slow-motion leading-order approximation of the\footnote{For a post-Newtonian, leading-order treatment including the PPN parameters $\beta$ and $\gamma$, see \citet{Bark76}. } General Theory of Relativity (GTR). In particular, we will analytically work out the long-term orbital effects experienced by an object of mass $m$ endowed with proper angular momentum $\bds\Sigma$ in  motion around a central body of mass $M$ and proper angular momentum $\bds J$. We will also generalize the \citet{LT} effect, caused by $\bds J$ on the orbital motion of a moving particle, assumed non-spinning, to arbitrary spatial orientations of it; {it may have important consequences from an empirical point of view when data from specific astronomical scenarios are analyzed}. The case of arbitrary masses $M_{\rm A}$ and $M_{\rm B}$ will be taken into account as well.

For a general review on the gravitational spin-dependent two-body problem, see \citet{Bark79}; see also \citet{Kri99} and \citet{Kri}.

According to \citet{Bark70}, the leading-order spin-spin and spin-orbit perturbations to the reduced Hamiltonian are, in the slow-motion limit and for {the extreme mass ratio case} $ M\gg m $,
\begin{equation}
\left\{
\begin{array}{lll}
{\mathcal{H}}_{\sigma}& = & \rp{G}{c^2 r^5}\left[3\left(\bds J\bds\cdot\bds r\right)\left(\bds\sigma\bds\cdot\bds r\right)-r^2\left(\bds\sigma\bds\cdot\bds J\right)\right], \\ \\
{\mathcal{H}}_{\sigma L}& = &\rp{3GM}{2c^2  r^3}\left(\bds\sigma\bds\cdot\bds L\right).
\end{array}
\right.\lb{hami}
\end{equation}
 In it $G$ is the Newtonian constant of gravitation and $c$ is the speed of light in vacuum. Moreover,  $\bds\sigma\doteq\bds\Sigma/m$ and $\bds L\doteq \bds r\bds\times\bds v$ are the reduced spin and orbital angular momentum of the particle moving with speed $v$ at distance $r$ from the central body: $[\sigma]=[L]=$ L$^2$ T$^{-1}$, so that $[\mathcal{H}_{\sigma}]=[\mathcal{H}_{\sigma L}]=$ L$^2$ T$^{-2}$, i.e. they have the dimensions of an energy per unit mass.
 {
 We point out that \rfr{hami}, and also \rfr{accas} treated later on to generalize the Lense-Thirring effect, are a direct consequence of the full spin-orbit Hamiltonian for two arbitrary masses \citep{Bark75}: see \rfr{parte} and \rfr{tutta}-\rfr{tutta2} in Section \ref{pulzar} for the connections between such scenarios.
 }

The spin-spin term ${\mathcal{H}}_{\sigma}$ was obtained for the first time by \citet{Schiff}, while the spin-orbit term\footnote{It is the gravitational analogue of the Thomas precession in electromagnetism which, curiously, was obtained 5 years later \citep{Thom}.} ${\mathcal{H}}_{\sigma L}$ is due to \citet{fok}; see also \citet{Bark70} for another derivation following a different line of reasoning. ${\mathcal{H}}_{\sigma }$ yields a   gravitational acceleration \textit{\`{a}} \textit{la} Stern-Gerlach \citep{Bark70,Bark75,Mash00,Chic05,Far06} whose effects on the orbital motion of a spinning particle were recently worked out in detail by \citet{Iorio} under certain simplifying assumptions about the orientation of $\bds J$ and the overall configuration of the system. For other studies concerning various consequences of the gravitational Stern-Gerlach effects on the orbital motion, see, e.g., \citet{Bark70,Bark75,Bark79,Far04,Bini1,Bini2,Chic05,Mash06,Far06}. Orbital effects of the spin-orbit term were derived by \citet{Bark70,Bark75}.
Concerning the role of general relativistic spin-spin interactions in Hawking
radiation from rotating black holes, see \citet{Haw}.


{Great efforts have been devoted so far to work out, with a variety of techniques, several kinds of next-to-leading order spin-spin and spin-orbit effects in compact binary systems for} {various ranges of} masses, spins and orbital geometries {in view of the impact that they would have  on the emission of gravitational waves; see the review by \citet{Blan}. We recall that, actually, several gravitational waves observatories are planned-and also already operating-to view binaries in various stages of their evolution \citep{Giaz,Fair}, so that they might detect them}. For two-body systems effects involving spins linearly\footnote{We mean that each spin enters the equations with the first power.}, see\footnote{{For example, the results by \citet{Konig} are valid for binary systems with equal masses and arbitrary spins, and arbitrary masses with only one of them spinning. \citet{Corn} derive a complete analytic solution for binary motion with spin-orbit interaction. }} \citet{Konig,nlo1,Tess,Tesspluri,Corn}. {See also \citet{Gerge1,Gerge2,Gerge3,Gerge4,Gerge5,Maja1,Maja2,Maja3} in which, among other things, different sets of independent variables were used. We also mention that \citet{nlo2} and \citet{tedeschi1} investigated next-to-leading-order spin-squared dynamics of general compact binaries, while \citet{tedeschi2} dealt with the next-to-leading order spin-orbit and spin-spin effects for a system of N gravitating spinning compact objects. In general, we will not deal in more details with such a host of gravitational wave-related features of motion since, as it will be clear \textit{a  posteriori} from our analysis itself, the current observational methods are quite insufficient to resolve such very peculiar spin-spin and spin-orbit effects.} \textcolor{black}{A direct and explicit comparison of our results with some of the leading order ones existing in literature would be difficult to be made in a meaningful way because of the differences in the approaches and the languages followed, driven by different original tasks. The tone of our work is mainly phenomenological, so that we choose the most widely adopted parameterization  in interpreting the empirical analyses usually performed in astronomical and astrophysical scenarios like the planetary motions in the solar system and  binary pulsars.}

The plan of the paper is as follows. In Section \ref{due} we  analytically work out the long-term temporal changes of order $\mathcal{O}(c^{-2})$ for all the Keplerian orbital elements of spinning and non-spinning particles orbiting a massive  central source.  Contrary to some other works on such topic existing in literature, we will not restrict ourselves to specific, simplified configurations: our results will be quite general,  as far as both the spatial orientations of the spins and the orbital geometry of the particle are concerned. In Section \ref{tre} we apply our results to \textcolor{black}{the aforementioned} specific astronomical and astrophysical scenarios.
Section \ref{quattro} summarizes our findings.
\section{Analytical calculation}\lb{due}
In order to evaluate the long-term effects caused by \rfr{hami}  on the  orbital motion of the moving  particle, it must, first, be evaluated onto the unperturbed\footnote{Notice that, in principle, it would be, perhaps, possible to use the perturbative techniques devised by \citet{Cal1,Cal2} for various first order post-Newtonian reference orbits involving non-spinning bodies, but they would yield additional negligible mixed orbital effects of order higher than $\mathcal{O}(c^{-2})$.} Keplerian orbit and averaged over one orbital revolution. {A similar approach can be found in \citet{DamDe}.}
The orbital motion in the  two-body problem can be usefully parameterized in terms of the standard osculating Keplerian orbital parameters \citep{Cap}. They are the semimajor axis $a$ and the eccentricity $e$, which fix the size and the shape\footnote{It can be $0\leq e <1$: for $e=0$ the ellipse reduces to a circle. } of the ellipse, and the longitude of the ascending node $\Om$, the argument of pericenter $\omega$ and the inclination $I$, which determine the orientation of the ellipse in the inertial space\footnote{They can be thought as the three Euler angles establishing the fixed orientation of a rigid body, i.e. the Keplerian ellipse which changes neither its size nor its shape, with respect to an inertial reference frame.}. The instantaneous position of the test particle along the Keplerian ellipse is reckoned by the true anomaly $f$, which, in combination with $\omega$, yields the argument of the latitude $u\doteq \omega + f$. The unperturbed Keplerian mean motion is $n\doteq \sqrt{GM/a^3}$; it is connected with the mean anomaly $\mathcal{M}$ through $\mathcal{M}\doteq n (t-t_{\rm p})$, where $t_{\rm p}$ is the time of passage at the pericenter. The Keplerian mean motion is also related to the orbital period $P_{\rm b}$ by $n\doteq 2\pi/P_{\rm b}$.

By recalling that on the unperturbed Keplerian ellipse it is \citep{Cap}
\begin{equation}
\left\{
\begin{array}{lll}
r & = & \rp{a(1-e^2)}{1+e\cos f}, \\ \\
x& = & r\left(\cos\Om\cos u-\cos I\sin\Om\sin u\right),\\ \\
y& = & r\left(\sin\Om\cos u+\cos I\cos\Om\sin u\right), \\ \\
z & = & r\sin I\sin u,\\ \\
dt & = & \rp{(1-e^2)^{3/2}}{n \left(1+e\cos f \right)^2}df,
\end{array}\lb{scazza}
\right.
\end{equation}
%
%
%
%
%
it turns out
\begin{equation}
\begin{array}{lll}
\left\langle {\mathcal{H}}_{\sigma}\right\rangle & = &\rp{G}{2c^2 a^3 (1-e^2)^{3/2}}\left\{  2\left(\bds\sigma\bds\cdot\bds J\right) +3\left(J_x \cos\Om +J_y \sin\Om\right)\left(\sigma_x \cos\Om+\sigma_y \sin\Om\right) +\right. \\ \\
 & + &\left. \left[J_z \sin I+ \cos I\left(J_y \cos\Om-J_x \sin\Om\right)\right]\left[\sigma_z \sin I + \cos I\left(\sigma_y \cos\Om-\sigma_x \sin\Om\right)\right]\right\},
\end{array}\lb{hamil1}
\end{equation}
{in which the brackets $\left\langle\cdots\right\rangle$ denote the average of their content over one orbital period.} {We computed it by using the true anomaly as independent variable:
\eqi \left\langle {\mathcal{H}}_{\sigma}\right\rangle=\rp{1}{P_{\rm b}}\int_0^{P_{\rm b}}{\mathcal{H}}_{\sigma} dt= \rp{n}{2\pi}\int_0^{2\pi}{\mathcal{H}}_{\sigma}(f)df\eqf
by means of \rfr{scazza}. Notice that, in principle, other averaging procedures implying the use of the mean anomaly $\mathcal{M}$ and the eccentric anomaly $E$ as independent variables may have been adopted; see, e.g., \citet{Gopa} for one based on the use of  $E$. It turns out that, in the present case, the use of $f$ through \rfr{scazza} allows to obtain exact expressions in $e$. Instead, it would not be possible with $\mathcal{M}$ and $E$ since necessarily approximated\footnote{{It is, e.g., $\cos f= \cos \mathcal{M}-e+e\cos 2\mathcal{M}+\mathcal{O}(e^2),\ \sin f= \sin\mathcal{M}+e\sin 2\mathcal{M}+\mathcal{O}(e^2)$. \textcolor{black}{For $\cos E$ and $\sin E$ as infinite summations of trigonometric functions over $\mathcal{M}$, see, e.g., \citet{TesScia}.}}} expressions for $\cos f$ and $\sin f$ or, equivalently, of $\cos E$ and $\sin E$, which enter $\mathcal{H}_{\sigma},\mathcal{H}_{\sigma L},\mathcal{H}_{JL}$,  in terms of $\mathcal{M}$ to some order in $e$ should be used \citep{Smart,Kova}.
}
The dimensions of $\left\langle {\mathcal{H}}_{\sigma}\right\rangle$  are correctly $[\left\langle \mathcal{H}_{\sigma}\right\rangle]=$ L$^2$ T$^{-2}$.
In the case of ${\mathcal{H}}_{\sigma L}$, it must be recalled that, in the unperturbed case, the reduced orbital angular momentum $\bds L$ is a vector directed along the out-of-plane direction, determined by the unit vector $\bds{\hat{\mathfrak{n}}}$ \citep{Sof}
\begin{equation}
\bds{\hat{\mathfrak{n}}}=\left\{
\begin{array}{lll}
\sin I\sin\Om, \\ \\
-\sin I\cos\Om, \\ \\
 \cos I,
\end{array}\lb{nu}
\right.
\end{equation}
and whose magnitude is \citep{Cap}
\eqi L = na^2\sqrt{1-e^2}.\eqf
Thus,
\eqi\left\langle {\mathcal{H}}_{\sigma L}\right\rangle = \rp{3GM n}{2c^2  a (1-e^2)} \left[\sigma_z \cos I +\sin I\left(\sigma_x \sin\Om -\sigma_y\cos\Om
\right)\right]:\lb{hamil2}\eqf also in this case, $[\left\langle \mathcal{H}_{\sigma L}\right\rangle]=$ L$^2$ T$^{-2}$, as expected.
Let us note that the results of \rfr{hamil1} and \rfr{hamil2} are exact in the sense that  approximations neither in $e$ nor in $I$ have been used to obtain them. Moreover, \rfr{hamil1} and \rfr{hamil2} are defined also for $e\rightarrow 0,\ I\rightarrow 0$.

It is, now, possible to straightforwardly work out
the secular variations of all the six Keplerian orbital elements through the Lagrange planetary equations. They are\footnote{Note that  the convention $U_g\doteq GM/r$, yielding $\bds F=\bds\nabla U_g$,  is often used  in many textbooks for the Newtonian gravitational potential $U_g$; in such a case, the right-hand-sides of the Lagrange planetary equations have a minus sign with respect to  \rfr{lagra}.  } \citep{BeFa}
\begin{equation}
\left\{
\begin{array}{lll}
  \rp{d a}{dt} &=& -\rp{1}{na}\left(2\rp{\partial \mathcal{R}}{\partial \mathcal{M}}\right), \\  \\
  \rp{d e}{dt} &=& \rp{1}{na^2}\left(\rp{1-e^2}{e}\right)\left( \rp{1}{\sqrt{1-e^2}}\rp{\partial \mathcal{R}}{\partial\omega} -  \rp{\partial \mathcal{R}}{\partial\mathcal{M}} \right), \\ \\
  \rp{d I}{dt} &=& \rp{1}{na^2\sqrt{1-e^2}\sin I}\left( \rp{\partial \mathcal{R}}{\partial\Om} -\cos I  \rp{\partial \mathcal{R}}{\partial\omega} \right), \\ \\
  \rp{d \Om}{dt} &=& -\rp{1}{na^2\sqrt{1-e^2}\sin I}\left( \rp{\partial \mathcal{R}}{\partial I}  \right), \\ \\
  \rp{d \varpi}{dt} &=& -\rp{1}{na^2}\left [ \rp{\sqrt{1-e^2}}{e}\rp{\partial \mathcal{R}}{\partial e} +\rp{\tan (I/2)}{\sqrt{1-e^2}}  \rp{\partial \mathcal{R}}{\partial I} \right], \\ \\
  \rp{d \mathcal{M}}{dt} &=& n + \rp{1}{na^2}\left[ 2a\rp{\partial \mathcal{R}}{\partial a} +\left(\rp{1-e^2}{e}\right) \rp{\partial \mathcal{R}}{\partial e} \right],
  \end{array}
\right.\lb{lagra}
\end{equation}
in which $\varpi\doteq \Om+\omega$ is the longitude of the pericentre, one of the parameters usually estimated by the astronomers when they process long data records of the planets of our solar system, and $\mathcal{R} $ denotes the averaged perturbation of the Newtonian gravitational potential: in our case it is
\eqi \mathcal{R} = \left\langle {\mathcal{H}}_{\sigma}\right\rangle + \left\langle {\mathcal{H}}_{\sigma L} \right\rangle. \eqf
A rapid inspection of \rfr{hamil1} and \rfr{hamil2} tells us that, according to \rfr{lagra}, all the Keplerian orbital elements should, in principle, experience secular variations, apart from the semi-major axis $a$.
\subsection{The spin-spin Stern-Gerlach  precessions}
Actually, for ${\mathcal{H}}_{\sigma}$ of \rfr{hamil1} the secular changes are
\begin{equation}
\left\{
\begin{array}{lll}
\dert a t & = & 0, \\ \\
\dert e t & = & 0, \\ \\
\dert I t & = & \rp{3 G}{2 c^2 a^5 n (1-e^2)^2}\left\{  \sin I\left[\left(J_y\sigma_x + J_x\sigma_y \right)\cos 2\Om + \left(J_y\sigma_y -J_x\sigma_x\right)\sin 2\Om\right]  - \right. \\ \\
&-&\left. \cos I \left[ \left(J_z\sigma_x + J_x\sigma_z\right)\cos\Om + \left(J_z\sigma_y + J_y\sigma_z\right)\sin\Om\right]\right\}, \\ \\
\dert\Om t & = & \rp{3G}{2c^2 a^5 n (1-e^2)^2}\left\{\cos 2 I\csc I\left[\left(J_z\sigma_x + J_x\sigma_z\right)\sin\Om -\left(J_z\sigma_y + J_y\sigma_z\right)\cos\Om \right]+\right.\\ \\
&+&\left. \cos I\left[\bds J\bds\cdot\bds\sigma-3J_z\sigma_z +\left(J_y\sigma_y-J_x\sigma_x\right)\cos 2\Om-\left(J_x\sigma_y+J_y\sigma_x\right)\sin 2\Om\right]\right\},\\ \\
\dert\varpi t & = &
\rp{3G}{2c^2  a^5 n(1-e^2)^2}
\left\{
2\bds J\bds\cdot\bds\sigma -3\left(J_x \cos\Om + J_y \sin\Om\right) \left(\sigma_x \cos\Om + \sigma_y \sin\Om\right)-\right. \\ \\
&-&\left.  3\left[J_z \sin I+ \cos I \left(J_y\cos\Om -J_x \sin\Om\right)\right] \left[\sigma_z \sin I+ \cos I \left(\sigma_y \cos\Om - \sigma_x \sin\Om\right)\right] -\right. \\ \\
&-&\left.
\tan\left(I/2\right)\left[J_z \cos I +
      \sin I\left(-J_y \cos\Om +
            J_x \sin\Om\right)\right] \left[\sigma_z \sin I+
      \cos I \left(\sigma_y \cos\Om - \sigma_x \sin\Om\right)\right]-\right. \\ \\
&-&\left.
\tan\left(I/2\right)\left[J_z \sin I+
      \cos I \left(J_y \cos\Om -
            J_x \sin\Om\right)\right]
      \left[\sigma_z \cos I +
      \sin I\left(-\sigma_y \cos\Om + \sigma_x \sin\Om\right)\right]
 \right\}, \\ \\
\dert{\mathcal{M}} t & = &-\rp{3G}{2c^2 a^5 n(1-e^2)^{3/2}}\left\{ -2\left(\bds J\bds\cdot\bds\sigma\right) +3\left(J_x\cos\Om + J_y\sin\Om\right)\left(\sigma_x\cos\Om+\sigma_y\sin\Om\right)  +\right. \\ \\
& + &\left. 3\left[J_z\sin I+\cos I\left(J_y\cos\Om-J_x\sin\Om\right)\right] \left[\sigma_z\sin I+\cos I\left(\sigma_y\cos\Om-\sigma_x\sin\Om\right)\right]\right\}.
\end{array}\lb{grosso}
\right.
\end{equation}
Their dimensions are correctly those of T$^{-1}$.
%
\subsection{The spin-orbit precessions}
\subsubsection{The case of a small spinning object orbiting around a non-rotating central body}
The precessions caused by ${\mathcal{H}}_{\sigma L}$ in \rfr{hamil2} are simpler, amounting to
\begin{equation}
\left\{
\begin{array}{lll}
\dert a t & = & 0, \\ \\
\dert e t & = & 0, \\ \\
\dert I t & = & \rp{3GM\left(\sigma_x \cos\Om+\sigma_y\sin\Om\right)}{2c^2 a^3(1-e^2)^{3/2}}, \\ \\
\dert\Om t & = & \rp{3GM\left[ \sigma_z + \cot I\left(\sigma_y\cos\Om -\sigma_x \sin\Om\right)\right]}{2c^2 a^3(1-e^2)^{3/2}}, \\ \\
\dert\varpi t & = & -\rp{ 3GM \left\{2\left[\sigma_z\cos I+\sin I\left(\sigma_x\sin\Om-\sigma_y\cos\Om \right)  \right]
-\left[\sigma_z\sin I+\cos I\left(\sigma_y\cos\Om-\sigma_x\sin\Om \right) \right]\tan(I/2)
\right\}}{2c^2 a^3(1-e^2)^{3/2}}, \\ \\
\dert{\mathcal{M}} t& = & 0;
\end{array}\lb{piccolo}
\right.
\end{equation}
also in this case, it is easily recognized that their dimensions are those of T$^{-1}$, as expected.
The results of \rfr{grosso} and \rfr{piccolo} are completely general, and are exact both in $e$ and in $I$. Moreover, they are valid also for $e\rightarrow 0,\ I\rightarrow 0$, apart from the node precessions which become singular for $I=0$. They may be compared with those obtained by \citet{Bark70}; concerning the occurrence of the change in the inclination, it was predicted by \citet{Bark70}.
\subsubsection{The case of a non-spinning particle orbiting around a rotating central body: a generalization of the Lense-Thirring effect}
Another larger term should, actually, be considered in addition to ${\mathcal{H}}_{\sigma L}$ in \rfr{hami}. It comes from the full two-body spin-orbit Hamiltonian \citep{Bark75}; in the {extreme mass ratio} limit $M\gg m$, it can straightforwardly be obtained from \rfr{hami} with the replacement
\eqi \rp{3}{2}M\bds\sigma\rightarrow 2\bds J.\lb{sos}\eqf Thus, we have
\eqi {\mathcal{H}}_{JL} = \rp{2G}{c^2 r^3}\left(\bds J\bds\cdot\bds L\right):\lb{accas}\eqf it is $[\mathcal{H}_{JL}]=$ L$^2$ T$^{-2}$. It can easily be recognized as just the gravitomagnetic Lense-Thirring term\footnote{See also \citet{Lan} and \citet{Ash}.} \citep{Iorio01}
\eqi \mathcal{H}_{JL}=-\mathcal{L}_{\rm gm}=\rp{\bds A_{\rm g}\bds\cdot \bds v}{c},\eqf
with the gravitomagnetic vector potential given by \citep{Mash07}
\eqi
\bds A_{\rm g}=\rp{2G\bds J\bds\times\bds r}{c r^3};
\eqf
$[A_{\rm g}]=$ L$^2$ T$^{-2}$.
Then, \rfr{piccolo}, with the replacement of \rfr{sos}, straightforwardly gives us the  generalized Lense-Thirring precessions of the Keplerian orbital elements of a test particle for an arbitrary orientation of the proper angular momentum of the source. They are
\begin{equation}
\left\{
\begin{array}{lll}
\dert a t & = & 0, \\ \\
\dert e t & = & 0, \\ \\
\dert I t & = & \rp{2G\left(J_x \cos\Om+J_y\sin\Om\right)}{c^2 a^3(1-e^2)^{3/2}}, \\ \\
\dert\Om t & = & \rp{2G\left[ J_z + \cot I\left(J_y\cos\Om -J_x \sin\Om\right)\right]}{c^2 a^3(1-e^2)^{3/2}}, \\ \\
\dert\varpi t & = & -\rp{ 2G \left\{2\left[J_z\cos I+\sin I\left(J_x\sin\Om-J_y\cos\Om \right)  \right]
-\left[J_z\sin I+\cos I\left(J_y\cos\Om-J_x\sin\Om \right) \right]\tan(I/2)
\right\}}{c^2 a^3(1-e^2)^{3/2}}, \\ \\
\dert{\mathcal{M}} t& = & 0;
\end{array}\lb{piccololt}
\right.
\end{equation}
their dimensions are  correctly those of T$^{-1}$.
On the contrary, all the usual derivations of the \citet{LT} precessions of the Keplerian orbital elements existing in literature are based on the particular choice of aligning $\bds J$ along the $z$ axis. To this aim, it can be noted that \rfr{piccololt}  yields\footnote{Indeed, by posing $I/2\doteq \alpha$, it is easy to show that $-2\cos I + \sin I\tan (I/2)=1-3\cos I$.} just the usual Lense-Thirring rates for $J_x,J_y\rightarrow 0$.  \rfr{piccololt} tells us that also the inclination $I$ experiences a secular gravitomagnetic change if $\bds J$ has components along the $x$ and $y$ axes as well: it is independent of the inclination $I$ itself. Moreover, $J_x$ and $J_y$ induce additional precessions for the node $\Om$ and the longitude of pericentre $\varpi$ which depend on $I$. In general, all the additional precessions depend on $\Om$ as well.
\section{Observability of the computed effects}\lb{tre}
In this Section, we compare our predicted results to the latest observationally determined orbital precessions in some astronomical and astrophysical scenarios. We will work them in detail, also because the results obtained  here for the spin configurations may turn out to be useful also for different dynamical effects.
\subsection{The Sun-Mercury system}\lb{solsys}
Let us, first, apply our results to the solar system. In this case, the inclination $I$ may typically be referred, e.g., to  the ecliptic plane, assumed as  reference $\{xy\}$ plane, and the longitude of the ascending node $\Om$ is counted from the Vernal equinox, assumed directed along the positive $x$ axis.
Concerning the Sun's spin axis, its orientation with respect to the J2000 ecliptic is
\begin{equation}
\left\{
\begin{array}{lll}
\hat{J}^{\odot}_x & = & \cos\delta_0^{\odot}\cos\alpha_0^{\odot}= 0.122, \\ \\
\hat{J}^{\odot}_y & = & \cos\delta_0^{\odot}\sin\alpha_0^{\odot}\cos\varepsilon +\sin\delta_0^{\odot}\sin\varepsilon= -0.031,  \\ \\
\hat{J}^{\odot}_z & = & -\cos\delta_0^{\odot}\sin\alpha_0^{\odot}\sin\varepsilon +\sin\delta_0^{\odot}\cos\varepsilon = 0.992,
\end{array}
\right.
\end{equation}
where
\begin{equation}
\left\{
\begin{array}{lll}
\alpha_0^{\odot} &=& 286.13\ {\rm deg},\\ \\
\delta_0^{\odot} & = & 63.87\ {\rm deg},
\end{array}
\right.
\end{equation}
are the\footnote{The right ascension $\alpha$ and the declination $\delta$ refer to the mean terrestrial equator at J2000.} right ascension and declination of the Sun's north pole of rotation, respectively \citep{seidel}, while \eqi\varepsilon=23.439\  {\rm deg}\eqf is the obliquity of the Earth's equator to the ecliptic at J2000 \citep{Fuku}.
The magnitude of the Sun's proper angular momentum was measured with the helioseismology technique. It amounts to \citep{Pij}
\eqi J_{\odot}=  (190.0\pm 1.5)\times 10^{39}\ {\rm kg\ m^2\ s^{-1}}.\eqf

Instead, for Mercury we have
\begin{equation}
\left\{
\begin{array}{lll}
\hat{\sigma}^{\mercury}_x & = & \cos\delta_0^{\mercury}\cos\alpha_0^{\mercury}= 0.091, \\ \\
\hat{\sigma}^{\mercury}_y & = & =\cos\delta_0^{\mercury}\sin\alpha_0^{\mercury}\cos\varepsilon +\sin\delta_0^{\mercury}\sin\varepsilon= -0.081,  \\ \\
\hat{\sigma}^{\mercury}_z & = & -\cos\delta_0^{\mercury}\sin\alpha_0^{\mercury}\sin\varepsilon +\sin\delta_0^{\mercury}\cos\varepsilon = 0.993,
\end{array}
\right.
\end{equation}
from \citep{seidel}
\begin{equation}
\left\{
\begin{array}{lll}
\alpha_0^{\mercury} &=& 281.01\ {\rm deg},\\ \\
\delta_0^{\mercury} & = & 61.45\ {\rm deg},
\end{array}
\right.
\end{equation}
and
\eqi \sigma_{\mercury} = 2.4\times 10^6\ {\rm m^2\ s^{-1}},\eqf as can be inferred from the values of its equatorial radius, angular rotation speed and normalized polar moment of inertia\footnote{See also http://nssdc.gsfc.nasa.gov/planetary/factsheet/mercuryfact.html on the WEB.} \citep{Thol,seidel}.

Since \citep{Murr}
\begin{equation}
\left\{
\begin{array}{lll}
a_{\mercury} & = &0.387\ {\rm AU}, \\ \\
e_{\mercury} & = & 0.205, \\ \\
I_{\mercury} & = & 7.005\ {\rm deg},\\ \\
\Om_{\mercury}  &=& 48.330\ {\rm deg},
\end{array}
\right.
\end{equation}
it turns out that the spin-spin and spin-orbit precessions of Mercury are up to $5-11$ orders of magnitude smaller than the present-day level of accuracy in measuring them \citep{Pit}, i.e. about a few milliarcseconds per century (mas cty$^{-1}$ in the following).
Indeed, \rfr{grosso} yields for the ${\mathcal{H}}_\sigma-$related precessions
\begin{equation}
\left\{
\begin{array}{lll}
\dert I t & = & -4\times 10^{-11}\ {\rm mas\ cty^{-1}},\\ \\
\dert \Om t & = &  -5\times 10^{-11}\ {\rm mas\ cty^{-1}}, \\ \\
\dert\varpi t & = & 1.35\times 10^{-9}\ {\rm mas\ cty^{-1}}, \\ \\
\dert{\mathcal{M}} t & = & 1.33\times 10^{-9}\ {\rm mas\ cty^{-1}},
\end{array}
\right.
\end{equation}
while the Mercury-induced spin-orbit rates of \rfr{piccolo} are
\begin{equation}
\left\{
\begin{array}{lll}
\dert I t & = & 3\times 10^{-9}\ {\rm mas\ cty^{-1}}, \\ \\
\dert\Om t & = & -1.2\times 10^{-8}\ {\rm mas\ cty^{-1}}, \\ \\
\dert \varpi t & = & -3.8587\times 10^{-5}\ {\rm mas\ cty^{-1}}.
\end{array}
\right.
\end{equation}
From the extremely small values of the secular rates of the inclination it can be inferred that the general relativistic spin-spin and spin-orbit effects considered here did not play any role in the solar system's evolution throughout its lifetime of about $4.5$ Gyr.

Moving to the  Lense-Thirring precessions of \rfr{piccololt}, they are
\begin{equation}
\left\{
\begin{array}{lll}
\dert I t & = & 0.06\ {\rm mas\ cty^{-1}}, \\ \\
\dert\Om t & = & 0.08\ {\rm mas\ cty^{-1}}, \\ \\
\dert \varpi t & = & -2.01\ {\rm mas\ cty^{-1}}.
\end{array}
\right.\lb{terremoto}
\end{equation}
The secular rate of the inclination of the orbit of Mercury is, at present, too small by likely two orders of magnitude to be measurable in the near future. An important result for the node is that the additional precession due to $J^{\odot}_x$ and $J^{\odot}_y$, as large as $-0.92$ mas cty$^{-1}$,  almost cancels the precession caused only by $J_z$, which amounts to\footnote{\citet{Cug78} gave a value too large by one order of magnitude.} 1 mas cty$^{-1}$ \citep{Iorio05}. Thus, the net Lense-Thirring node precession is  two orders of magnitude smaller than the  present-day accuracy in determining the nodal rates for the inner planets. On the contrary, the total precession of $\varpi$  is practically equal to the value\footnote{\citet{Cug78} and \citet{Sof} released values  larger by a factor $10-5$, respectively, while the figure by \citet{Bark70} was only $1.5$ times larger.} $-1.98$ mas cty$^{-1}$ obtained by assuming $J_x=J_y=0$ \citep{Iorio07}.
\textcolor{black}{A recent application of such results to the motion of Mercury in view of the latest observations from the MESSENGER spacecraft orbiting it can be found in \citet{Iorio011}.}

{
Finally, we point out that the still unmodelled (and undetected)  2PN pointlike perihelion precession \citep{Damscia,Wex}, proportional to $\dot\varpi_{\rm 2PN}\propto (GM)^{5/2}c^{-4}a^{-7/2}$ up to terms of order $\mathcal{O}(e^2)$, is about $10^{-3}$ mas cty$^{-1}$ for Mercury.
}
\subsection{The double pulsar PSR J0737-3039A/B system}\lb{pulzar}
A relatively more promising scenario is represented by the double pulsar  PSR J0737-3039A/B \citep{Bur,Lyne}: it is a close binary system {at about\footnote{\textcolor{black}{Actually, doubts on the reliability of such an estimate may be cast since it is based on a model of the interstellar electron density. On the other hand, subsequent very long baseline interferometry (VLBI) observations \citep{Deller} yield a distance of $1150_{-160}^{+220}$ pc.}} 500 pc from us \citep{Kra}} constituted of two neutron stars both visible as radiopulsars\footnote{Actually, B disappeared from our view in 2009 \citep{Pere}.} orbiting along moderately eccentric barycentric ellipses in  {$2.45$ hr$=0.102$ d \citep{Kra}}. In this case the inclination $I$ refers to the plane of the sky, assumed as reference $\{xy\}$ plane, so that the $z$ axis is directed towards the observer along the line-of-sight direction.
The relevant parameters of such a system are\footnote{{Concerning the inclination $I$, in the PSR J0737-3039A/B system it is, actually, inferred from the  determined value of $\sin I$ through the Shapiro delay measurements. Thus, in addition to the quoted value of $I=88.69$ deg, also the value $I=180$ deg$-88.69$ deg$=91.31$ deg is, in principle, admissible. From a practical, calculational point of view, this has no appreciable consequences on the numerical results presented below. }} \citep{Kra}
\begin{equation}
\left\{
\begin{array}{lll}
\mathfrak{M} & \doteq & M_{\rm A}+M_{\rm B} =2.58708\ {\rm M}_{\odot}, \\ \\
M_{\rm A} & = & 1.3381\ {\rm M}_{\odot}, \\ \\
M_{\rm B}& = & 1.2489\ {\rm M}_{\odot}, \\ \\
\mathfrak{r}& \doteq & \rp{M_{\rm A}}{M_{\rm B}} = 1.0714,\\ \\
\mu & \doteq & \rp{M_{\rm A} M_{\rm B}}{M_{\rm A} + M_{\rm B}} = 0.6459\ {\rm M}_{\odot}, \\ \\
\nu_{\rm A} & = & 44.05\ {\rm Hz}, \\ \\
\nu_{\rm B} & = & 0.36\ {\rm Hz}, \\ \\
a & = & 8.78949386\times 10^8\ {\rm m}, \\ \\
e &=& 0.0877775, \\ \\
I &=& 88.69\ {\rm deg}, \\ \\
\mu L & = & 7.03066\times 10^{44}\ {\rm kg\ m^2\ s^{-1}},\\ \\
J_{\rm A} & = & 2.768\times 10^{40}\ {\rm kg\ m^2\ s^{-1}}, \\ \\
\Sigma_{\rm B} & = & 2.3\times 10^{38}\ {\rm kg\ m^2\ s^{-1}},
\end{array}\lb{datipuls}
\right.
\end{equation}
where we used the standard value of the pulsar's moment of inertia $\mathcal{I}\approx  10^{38}$ kg m$^2$ and the fact that the rotational spin periods of A and B are $T_{\rm A}\doteq 1/\nu_{\rm A}=23$ ms \citep{Bur} and $T_{\rm B}\doteq 1/\nu_{\rm B}=2.8$ s \citep{Lyne}, respectively.
{From \rfr{datipuls} it can be noticed that the longitude of the ascending node $\Om$ does not figure in the list of known parameters of PSR J0737-3039A/B. It is so because $\Om$ does not directly affect the direct observable quantity, i.e. the standard timing formula \citep{timing1,timing2}, so that it cannot, in general, be determined from pulsars' timing data analysis. For a general review, see, e.g., \citet{psr}. The node could be measured only if a pulsar is close enough to the Earth. Indeed, in this case  the orbital motion of the Earth changes the apparent inclination angle $I$ of the pulsar orbit on the sky, an effect known as the annual-orbital parallax \citep{paralla}. It results in a periodic change of the projected semi-major axis. A second contribution comes from the transverse motion in the plane of the sky \citep{Kope}, yielding a secular variation of the projected semi-major axis. By including both these effects in the model of the pulse arrival times,  the longitude of the ascending node $\Om$ can be determined, as in the case of PSR J0437-4715 \citep{vanstra}, which is at only 140 pc from us.}

Concerning the orientations of the spin axes, it turns out that $\bds {J}_{\rm A}$ is substantially aligned\footnote{{Actually, according to \citet{Ferd}, there is still a $30\%$ chance that the tilt of A's spin is larger than 6 deg. This implies that the projection of $\bds J_{\rm A}$ on the orbital plane may be larger than $10\%$ of the total.}} with the total angular momentum of the system \citep{Man05,Ferd}, which practically coincides with the orbital angular momentum $\mu\bds L$, as can be inferred from \rfr{datipuls}. Thus, according to \rfr{nu}, we can pose
\begin{equation}
\left\{
\begin{array}{lll}
\hat{J}^{\rm A}_x & = & 0.997 \sin\Om, \\ \\
\hat{J}^{\rm A}_y & = & -0.997\cos\Om, \\ \\
\hat{J}^{\rm A}_z & = & 0.023:
\end{array}
\right.
\end{equation}
$\bds J_{\rm A}$ lies almost entirely in the plane of the sky.
Instead,  $\bds\sigma_{\rm B}$, which  undergoes the general relativistic de Sitter precession\footnote{It is the main candidate to explain the disappearance of the radiopulses of B in 2009; if it is the sole cause of such a phenomenon, the B's signal should reappear in about 2035 or even earlier, depending on the beam shape \citep{Kra10}.} \citep{Bark75,Boe75} describing a full cycle in 75 yr, has a different orientation \citep{Bret,Pere}. At the epoch May 2, 2006 (MJD 53857){, according to Figure 1 of \citet{Bret}}, it is
\begin{equation}
\left\{
\begin{array}{lll}
\hat{\sigma}_{x^{{'}}}^{\rm B}& = & \sin\phi\cos\theta = -0.501, \\ \\
\hat{\sigma}_{y^{{'}}}^{\rm B}& = & \sin\phi\sin\theta = 0.597, \\ \\
\hat{\sigma}_{z^{{'}}}^{\rm B}& = & \cos\phi = 0.625,
\end{array}\lb{siggy}
\right.
\end{equation}
where
\begin{equation}
\left\{
\begin{array}{lll}
\phi & = & 51.21 \ {\rm deg}, \\ \\
\theta & = & 130.02 \ {\rm deg}.
\end{array}
\right.
\end{equation}
{Actually, the frame used by \citet{Bret},  dubbed $\{x^{'},y^{'},z^{'}\}$ by us, is different from the one $\{x,y,z\}$ used here. Indeed, the $x^{'}$ axis is directed along the line-of-sight towards the Earth,  the $z^{'}$ axis is  directed along the (projected) orbital angular momentum, i.e. it is $\bds{\hat{z}}^{'}=\sin I\bds{\hat{\mathfrak{n}}}$, and the $y^{'}$ axis coincides with the line of the nodes. In other words, the $y^{'}$ and the $z^{'}$ axes lie in the plane of the sky, being rotated by $\Om$ with respect to the $x$ and $y$ axes. Thus, $\hat{\sigma}_x^{\rm B},\hat{\sigma}_y^{\rm B}$ are linear combinations\footnote{{They are $\hat{\sigma}^{\rm B}_x=\cos\Om\hat{\sigma}^{\rm B}_{y^{'}}+\sin^2 I\sin\Om\hat{\sigma}^{\rm B}_{z^{'}}$ and $\hat{\sigma}^{\rm B}_y=\sin\Om\hat{\sigma}^{\rm B}_{y^{'}}-\sin^2 I\cos\Om\hat{\sigma}^{\rm B}_{z^{'}}$.}} of $\hat{\sigma}_{y^{'}}^{\rm B}$ and $\hat{\sigma}_{z^{'}}^{\rm B}$ involving $\Om$ as well, while $\hat{\sigma}_z^{\rm B}=\hat{\sigma}_{x^{'}}^{\rm B}$.}
%
%

Concerning the spin-spin and spin-orbit Hamiltonians of \rfr{hami}, they are to be modified in order to take into account the fact that the mass of B is almost equal to that of A. According to \citep{Bark75},  ${\mathcal{H}}_{\sigma}$ must simply be re-scaled by the dimensionless factor \eqi \rp{M_{\rm B}}{\mu}=1.933,\lb{parte}\eqf while in the spin-orbit term of \rfr{hami} the replacement \eqi \rp{3}{2}M_{\rm A}\rightarrow \rp{3}{2}M_{\rm A}\left(1+\rp{4}{3}\rp{M_{\rm B}}{M_{\rm A}}\right)\textcolor{black}{,\ {\rm A} \leftrightarrows {\rm B} }\eqf
must be done. Moreover, a second term containing the spin of A must be added, so that
\eqi {\mathcal{H}}_{\sigma L}+{\mathcal{H}}_{J L}=\rp{G}{c^2r^3}\left[\rp{3}{2}M_{\rm A}\left(1+\rp{4}{3}\rp{M_{\rm B}}{M_{\rm A}}\right)\left(\bds \sigma_{\rm B}\bds\cdot\bds {\textcolor{black}{L}}\right) +\rp{3}{2}M_{\rm B}\left(1+\rp{4}{3}\rp{M_{\rm A}}{M_{\rm B}}\right)\left(\bds {\textcolor{black}{j}}_{\rm A}\bds\cdot\bds {\textcolor{black}{L}}\right)\right]:\lb{tutta}\eqf
\textcolor{black}{here $\bds j_{\rm A}\doteq \bds J_{\rm A}/M_{\rm A}$. In order to facilitate a comparison with other works existing in literature like, e.g., \citet{Dam92} it is convenient to re-write \rfr{tutta}, after  a little algebra, as
\eqi {\mathcal{H}}_{\sigma L}+{\mathcal{H}}_{J L}=\rp{2G}{c^2r^3}\left[\left(1+\rp{3}{4}\rp{M_{\rm A}}{M_{\rm B}}\right)\left(\bds \Sigma_{\rm B}\bds\cdot\bds L\right) +\left(1+\rp{3}{4}\rp{M_{\rm B}}{M_{\rm A}}\right)\left(\bds J_{\rm A}\bds\cdot\bds L\right)\right].\lb{tutta2}\eqf Thus, the resulting orbital effects can straightforwardly be inferred from \rfr{piccololt} with the replacement
\eqi \rp{2G}{c^2}\rightarrow \rp{2G}{c^2}\left(1+\rp{3}{4}\rp{M_{\rm B}}{M_{\rm A}}\right)\lb{sosti}\eqf for the precession by $J_{\rm A}$ and vice-versa for the effects induced by $\bds\Sigma_{\rm B}$. Some of them can be compared with those worked out by \citet{Dam92}. For example, our expression for the precession of $I$ due to, say, $J_{\rm A}$ obtainable from \rfr{piccololt} with the substitution of \rfr{sosti} coincides with Eq. (3.27), written for Eq. (3.7), of \citet{Dam92}. Indeed, the unit vector $\bds{\rm I}\equiv\bds{\rm {i}}$ of the line of the nodes entering Eq. (3.27) of \citet{Dam92} has cartesian components\footnote{\textcolor{black}{See Eq. (2.12a) of \citet{Dam92}.}} ${\rm I}_x\equiv{\rm {i}}_x=\cos\Om, {\rm I}_y\equiv{\rm {i}}_y=\sin\Om,$ so that its scalar product with the the spin of A in Eq. (3.27) by \citet{Dam92} yields just the geometrical factor of our \rfr{piccololt}.
}

 As far as the  spin-obit precessions caused by $\bds J_{\rm A}$ \textcolor{black}{are} concerned, \rfr{nu} and \rfr{piccololt} tell us that, since we are assuming the spin axis of A exactly aligned with the orbital angular momentum, the precessions of $I$ and $\Om$ vanish, contrary to that of $\omega$ which becomes proportional to $\cos^2 I+\sin^2 I$: the dependence on $\Om$ disappears. A similar situation occurs for the Sun-Mercury system in which, as we have seen in Section \ref{solsys}, the Lense-Thirring precessions of $I$ and $\Om$ are two orders of magnitude smaller than the perihelion precession. The magnitude  of the periastron precession due to $\bds J_{\rm A}$ is \textcolor{black}{of the order of}
\eqi\dert\omega t \textcolor{black}{\simeq} - 3.7\times 10^{-4}\ {\rm deg\ yr^{-1}}; \eqf the present-day level of accuracy in  determining the periastron precession of the double pulsar from observations is $6.8\times 10^{-4}$ deg yr$^{-1}$ \citep{Kra}. {The issue of investigating the actual measurability of the spin-orbit periastron precession caused by $\bds J_{\rm A}$ was tackled in \citet{Iorio09,KraWex},} {along with the aliasing effects of the precessional effects caused by the 1PN and 2PN \citep{Damscia,Wex} non-spinning point particles contributions. Here we briefly recall that  the nominal value of the 2PN precession, not yet measured, is of the order of $10^{-4}$ deg yr$^{-1}$, with an uncertainty of $10^{-6}$ deg yr$^{-1}$ due to our imperfect knowledge of the system's total mass and semi-major axis \citep{Iorio09}. Instead, the uncertainty in the measured 1PN precession due to the errors in the system's parameters is of about $3\times 10^{-2}$ deg yr$^{-1}$ \citep{Iorio09}.}
{If one tries to take into account a possible tilt of $\bds J_{\rm A}$ with respect to $\bds{\hat{\mathfrak{n}}}$ \citep{Ferd}, the effects would be very small and undetectable. Indeed, \rfr{piccololt}\textcolor{black}{, with \rfr{sosti},} tells us that the non-zero secular variation of, say, the inclination would be
\eqi\dert I t =  \textcolor{black}{\left(\hat{J}_x^{\rm A}\cos\Om + \hat{J}_y^{\rm A}\sin\Om\right)1.8\times 10^{-4}\ {\rm deg\ yr}^{-1},}\eqf
where \textcolor{black}{it is intended that the components of $\bds{\hat{J}}_{\rm A}$ do not coincide with those} of $\bds{\hat{\mathfrak{n}}}$ given in \rfr{nu}. \textcolor{black}{Furthermore, $dI/dt$ is not an observable quantity in the timing observations of the double pulsar. Indeed, they are sensitive to a change in the projected semi-major axis \citep{Dam92}, which is proportional to $\sin I$. This fact, on the other hand, supports the previous argument  that the effect on $I$ is  negligible from an observational point of view since the rate of variation of the projected semi-major axis  is proportional to $\cos I$, which is small for the pulsar PSR J0737-3039A/B.}
 }
Instead, all the spin-orbit precessions due to $\bds\Sigma_{\rm B}$ do not, in general, vanish:
they are as large as up to $1-2\times 10^{-6}$ deg yr$^{-1}$.
The spin-spin Stern-Gerlach  rates are quite smaller, being all of the order of $2-4\times 10^{-11}$ deg yr$^{-1}$.
\section{Summary and conclusions}\lb{quattro}
In this paper, we  analytically worked out the secular variations of the six osculating Keplerian orbital elements of a spinning particle in free motion around a central, slowly rotating body caused by the general relativistic spin-spin and spin-orbit interactions of order $\mathcal{O}(c^{-2})$ in the weak-field and slow-motion approximation. We neither restricted to specific spatial orientations of the angular momenta involved nor to particular orbital configurations: the results obtained are, thus, exact with respect to such issues and quite general. We adopted the Lagrange planetary equations  after having averaged  the perturbing reduced Hamiltionians over one orbital revolution. As reference, unperturbed orbit we adopted the Keplerian ellipse: far smaller effects due to mixed static-spin general relativistic terms could have, in principle, been worked out, but we neglected them because of their negligible magnitude.

Concerning the spin-spin Stern-Gerlach-like term, it turns out that the inclination, the node, the longitude of pericenter and the mean anomaly undergo secular precessions depending on the particle's semi-major axis \textcolor{black}{as $a^{-5} n^{-1}\propto a^{-7/2}$} and the eccentricity. The dependence on the particle's inclination and node is rather complicated; all the components of both spins enter the formulas as well.

The spin-orbit term containing the particle's spin causes secular precessions of the inclination, the node and the longitude of pericenter which go as $a^{-3}$. The spin-orbit term containing the angular momentum of the central body, arising from the full two-body spin-orbit Hamiltonian, yields the usual Lense-Thirring effect for a specific orientation of the body's spin. Instead, for a generic orientation of it we obtained secular precessions for the inclination, the node and the longitude of pericenter which depend on all the components of the angular momentum of the central body, and on the node and the inclination of the orbital plane. The dependence on the semi-major axis and the eccentricity is as for the other spin-orbit term.

We generalized the results obtained to a generic two-body system with arbitrary masses.

We applied our predictions to some specific astronomical scenarios like the Sun-Mercury system and the double pulsar. Concerning Mercury, it turns out that
the Stern-Gerlach spin-spin precessions are too small to be detected by $9-11$ orders of magnitude with respect to the present-day level of accuracy in empirically determining the planetary precessions, which is a few milliarcseconds per century. The precessions caused by the Mercury's spin are too small by $5-11$ orders of magnitude. In regard to the  Lense-Thirring effect, the slight misalignment of the Sun's equator  with respect to mean ecliptic at J2000, usually adopted in all the planetary data reductions performed by the most influential teams of astronomers, induces a tiny secular precession of the inclination of about $0.06$ milliarcseconds per century. Concerning the node, its secular precession turns out to be reduced to $0.08$ milliarcseconds per year with respect to the typical value of 1 milliarcseconds per year usually computed by neglecting the departure of the solar equator from the ecliptic, while the precession of the longitude of the perihelion is slightly augmented to $-2.01$ milliarcseconds per year with respect to the standard value of $-1.98$ milliarcseconds per year. Such results are important in view of a possible reliable detection of the solar Lense-Thirring effect in the near future.

Moving to the double pulsar scenario, the Stern-Gerlach spin-spin precessions are about 8 orders of magnitude smaller than the present-day accuracy of $6.4\times 10^{-4}$ degrees per year in determining the orbital precessions from timing data. The spin-orbit effects due to the angular momentum of B, which has the slowest rotation in the pair, are too small by  2 orders of magnitude. Instead, the  spin-orbit precessions caused by the spin of A are all vanishing because of its alignment with the system's orbital angular momentum, apart from the periastron which precesses at a rate of $-3.7\times 10^{-4}$ degrees per year.

The full generality of our calculations with respect to the orbital and the spin spatial configurations allows them to be straightforwardly extended to other systems like extrasolar planets. Indeed, they present a rich variety of orbital configurations and spins orientations which make them potentially interesting candidates to apply our results, especially those regarding the generalized Lense-Thirring precessions. This may be the subject of a forthcoming paper.
\section*{Acknowledgements}
I am grateful to two anonymous competent referees for their pertinent critical remarks which greatly contributed to improve the manuscript.


\end{document}